\documentclass[epj]{svjour}

\usepackage{graphicx}
\usepackage{fancyhdr}
\def\makeheadbox{{
 }}
\def\mytitle{My title} 
\def\myauthors{My name}  
\def\mytype{My type of session}
\def\mysession{My session}

\def\mytitle{SUSY charged Higgs at the TeVatron.} 
\def\myauthors{G\'erald Grenier}    
\def\mytype{Contributed Talk}    
\def\mysession{Colliders - Higgs Phenomenology}

\begin{document}
\title{Search for supersymmetric charged Higgs bosons at the TeVatron.}
\author{G\'erald Grenier\inst{1}
\thanks{\emph{Email:grenier@ipnl.in2p3.fr}}%
\thanks{on behalf of CDF and {D\O} collaborations}%
}                     
%
%
\institute{Institut de Physique Nucl\'eaire de Lyon, CNRS/IN2P3, Universit\'e Lyon 1, Universit\'e de Lyon, F69622 Villeurbanne, France}
%
\date{}
\abstract{
The data collected at the TeVatron RunIIa have been used to look for supersymmetric
charged Higgs boson and Left-Right suspersymmetric doubly charged Higgs boson. 
No signal of such bosons has been found and this note reports on the current analyses
and their observed excluded domains in models parameter space.  
} 
\maketitle
\section{Introduction}
\label{sec:intro}
The analyses reported in this note cover searches for charged and doubly
charged Higgs bosons. The charged Higgs boson is a needed element of the 
Higgs sector of any supersymmetric model\cite{Nilles:1983ge}. 
The doubly charged Higgs is here 
looked for in the context of Left-Right supersymmetric theories\cite{Gunion:1989in}. 
The data samples used for the analyses have been collected by 
either the {D\O} and CDF detectors during the RunIIa of the 
TeVatron which collides proton and antiproton with a center-of-mass
energy of 1.96~TeV. Depending on the analysis, the integrated luminosity used ranges
from 192~pb$^{-1}$ to 1.1~fb$^{-1}$. The acceptances in pseudo-rapidity
for both the CDF and {D\O} detectors are shown in table~\ref{tab:etaaccept}.
%
\begin{table}
\caption{Pseudo-rapidity acceptances for various reconstructed 
objects for the {D\O} and CDF detectors.
\label{tab:etaaccept} }      
\begin{tabular}{lll}
\hline\noalign{\smallskip}
  & CDF & {D\O}   \\
\noalign{\smallskip}\hline\noalign{\smallskip}
electron & $|\eta|<2.0$ & $|\eta|<3.0$ \\
muon & $|\eta|<1.5$ & $|\eta|<2.0$ \\
muon trigger & $|\eta|<1.0$ & $|\eta|<2.0$ \\
precision tracking & $|\eta|<2.0$ & $|\eta|<3.0$ \\
jets & $|\eta|<3.6$ & $|\eta|<4.2$ \\
\noalign{\smallskip}\hline
\end{tabular}
\end{table}
Those pseudo-rapidities are implicitely used in all the analyses described
in this note.

\section{MSSM charged Higgs}
\label{sec:chargedHiggs}
The Minimal Supersymmetric Standard Model\cite{Nilles:1983ge} requires a Higgs
mechanism with two Higgs doublet leading to the existence of a  
charged Higgs boson.
At the TeVatron, the direct production of 
$\mathrm{p \bar{p} \to H^+ H^-}$ has a cross-section too low to provide
sensitivity for the discovery of charged Higgs bosons \cite{Carena:2002es}.
Therefore, The charged Higgs is looked for in the decay of the top quark. Throughout
this section, the signal searched for is \newline $\mathrm{p \bar{p} \to t \bar{t}+X}$ 
with one top decaying into $\mathrm{H^+ b}$ and the other into 
$\mathrm{b l \nu }$ where $l$ is either an electron or a muon.

\subsection{lepton+tau searches}
\label{subsec:Hplustotaunu}  The CDF collaboration reports a search for the signal 
of Sect.~\ref{sec:chargedHiggs} in the case where the charged Higgs decays 
exclusively into a tau and a neutrino\cite{CDF:8353}. In this analysis,
the tau is reconstructed only through its hadronic decays. The CDF tau
identification uses a likelihood based on variables characterizing a tau \cite{CDF:8353}.
Examples of such variables are isolation criteria in the tracking and the calo\-ri\-meter
or ratios of the transverse momentum of the tracks composing the 
reconstructed tau candidates. Four different likelihoods are used depending on the 
hadronic decay of the tau. The four likelihoods cover the following four kinds of
tau decay topology : (1) one track and no $\pi^0$, (2) one  
track and one or more $\pi^0$, (3) three tracks with or without $\pi^0$ and 
(4) two tracks with or without $\pi^0$. The likelihoods aim at separating 
hadronically decaying taus from jets. A tau candidate is identified as a hadronically 
decaying tau if the likelihood is above 0.65.

The analysis requires at least one electron or muon with a transverse momentum greater than
20 GeV/c, at least two jets with a transverse momentum greater than 25 GeV/c for the most
energetic one and greater than 15 GeV/c for the second most energetic one. At least one 
jet should be identified as a b-jet using a b-jet identification algorithm based on
displaced vertices. The missing transverse energy should be greater than 20 GeV. $H_T$,
the scalar sum of the transverse energy of all calorimeter objects should be greater 
than 205 GeV and the event should contain exactly one recontructed tau lepton.

\begin{figure}
\includegraphics[width=0.45\textwidth,height=0.25\textwidth,angle=0]{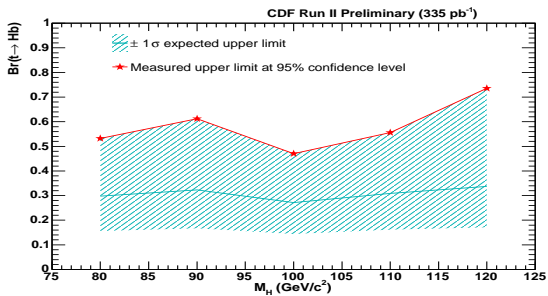}
\caption{95 \% confidence level upper limit on the branching fraction of 
$\mathrm{t \to b H^+}$ as a function of the charged Higgs mass.
\label{fig:cdfHtaunulimit} }       
\end{figure}

The observed numbers of events in a data sample of 335~pb$^{-1}$ are 4 for 
$1.90 \pm 0.26$ expected from the background in the electron+tau lepton channel and
2 for $1.97 \pm 0.27 $ expected in the muon+tau lepton channel.
Observed and expected numbers of events are in agreement. 
An upper limit on the branching fraction of $\mathrm{t \to b H^+}$
($\mathrm{Br(t \to b H^+)}$) at the 95\% confidence level
is derived assuming a 100 \% decay rate of
$\mathrm{H^+ \to \bar\tau \nu}$ and a Standard Model cross section of 
$\mathrm{p \bar{p} \to t \bar{t}+X}$ of 7.0~pb.
The limit is shown as a function of the 
charged Higgs mass on figure~\ref{fig:cdfHtaunulimit}. The red (resp. blue) line is the 
limit obtained by (resp. without) taking into account the background estimation
uncertainties.

\subsection{Di-top analyses reinterpretation} 
\label{subsec:Hplusinttbar}
If the decay $\mathrm{t \to H^+ b}$ is opened, then it will compete with the Standard
Model decay $\mathrm{t \to W^+ b}$. H$^+$ and W$^+$ can decay into similar channel 
but with different branching ratios. The CDF collaboration reports a reuse of its previous 
analyses used in the measurement of the $\mathrm{t \bar{t}}$ cross section to look for
possible signs of decay of H$^+$ \cite{CDF:Hpp}. The reused analyses are :
\begin{itemize}
\item one muon or electron + one muon or electron, 
below referred to as the dilepton analysis \cite{Acosta:2004uw}.
\item one muon or electron + jets with exactly one jet identified as a b-jet, 
below referred to as the lepton+ jets(1 tag) analysis \cite{Acosta:2004hw}.
\item one muon or electron + jets with two or more jets identified as b-jets, 
below referred to as the lepton+ jets(2 tags) analysis \cite{Acosta:2004hw}.
\item one muon or electron + one hadronically decaying tau, 
below referred to as the lepton+tau analysis \cite{Abulencia:2005et}.
\end{itemize}
The original analyses, done with a data sample corresponding to 192~pb$^{-1}$,  
are slightly modified to ensure that any selected event is selected exclusively by one 
of the four analyses. 
In the reinterpretation of the four analyses, the signal described at the beginning 
of section~\ref{sec:chargedHiggs} is looked for assuming that the top quark decays only
in W$^+$b and H$^+$b and that the H$^+$ can only decay in the following four final
states : $\mathrm{t^* \bar{b}}$ , $\mathrm{c \bar{s}}$ , $\bar\tau \nu_\tau$ , 
$\mathrm{W^+ h^0}$. Here the 
h$^0$ is reconstructed only through its decay in $\mathrm{b \bar{b}}$. If the H$^+$
decays exclusively in  $\mathrm{\bar\tau \nu_\tau}$ then the lepton+tau analysis will 
have an excess of events with respect to the Standard Model expectations while the 
dilepton and lepton+jets analyses will have a deficit of event. 

\begin{figure}
\includegraphics[width=0.45\textwidth,height=0.40\textwidth,angle=0]{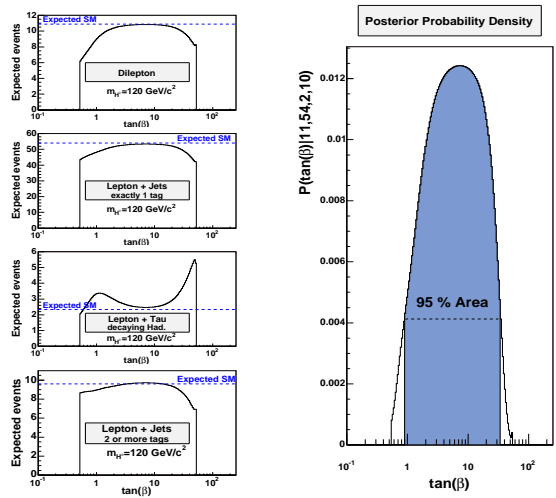}
\caption{Left: expected number of events to be seen by each of the four analyses 
for the Standard Model only and Standard Model+120 GeV/c$^2$ charged Higgs as a function 
of tan $\beta$. Right : posterior probability on tan $\beta$ (see text) assuming
the number of observed events equals the Standard Model expectation shown on the left. 
The region outside the blue band would be excluded.
\label{fig:ttprinciple} }      
\end{figure}

\begin{figure}
\includegraphics[width=0.45\textwidth,height=0.30\textwidth,angle=0]{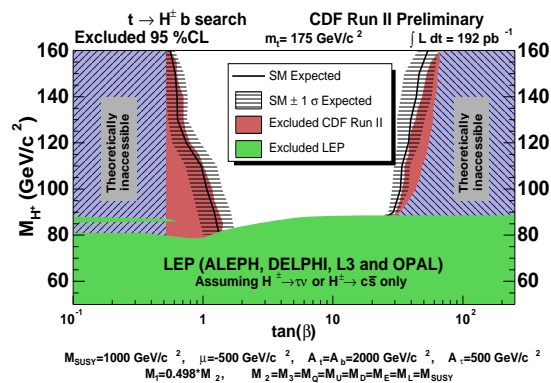}
\caption{Observed (red) and expected (black line) 95 \% confidence level exclusion  
region in the charged Higgs mass-tan $\beta$ plane for MSSM parameters displayed 
on the figure. 
\label{fig:ttresult1} }       
\end{figure}

For the reinterpretation, one MSSM parameter is chosen, for example tan $\beta$. 
The number of events expected for each of the four analyses is computed for the 
Standard Model and for the Standard Model+MSSM as a function of the MSSM parameter.
Such expectations are shown on the left of Figure~\ref {fig:ttprinciple} for a 
charged Higgs mass of 120 GeV/c$^2$ and a set of MSSM parameters described on 
figure~\ref{fig:ttresult1}. From these estimation, a likelihood based on Poisson
probability to observe N events is derived for each analyses. The four likelihoods,
the number of observed events and a prior probability on the MSSM parameter are
combined to derived a posterior probability as a function of the MSSM parameter.
This posterior probability is then used to exclude some values of the MSSM parameter
as illustrated on the right side of figure~\ref{fig:ttprinciple} for tan~$\beta$.

\begin{figure}
\includegraphics[width=0.45\textwidth,height=0.30\textwidth,angle=0]{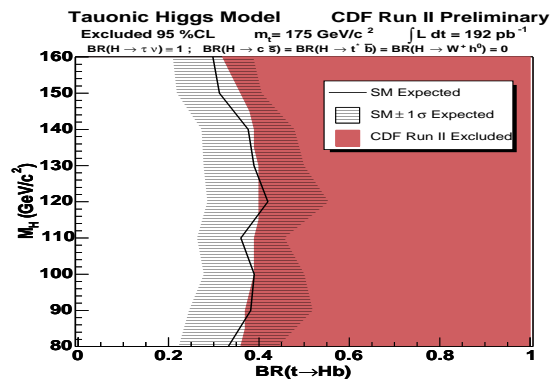}
\caption{95 \% confidence level excluded region in the charged Higgs mass - 
$\mathrm{Br(t \to H^+ b)}$ plane assuming $\mathrm{Br(H^+ \to \bar \tau \nu_\tau)=1}$. 
\label{fig:ttresult2} }      
\end{figure}

The observed and expected number of events are displayed in table~\ref{tab:cdftt}.
From there, excluded regions in the charged Higgs mass - tan~$\beta$ plane are derived
as shown on figure~\ref{fig:ttresult1}. On this figure, the red area is excluded by
the present analysis, the dark line is the Standard Model expected excluded area with its 
one standard deviation shown as horizontal black lines.
\begin{table}
\caption{Standard Model (SM) expected and observed number of events 
for the $\mathrm{t \bar{t}}$ analyses. The $\mathrm{t \bar{t}}$ 
cross section is assumed to be $6.7 \pm 0.9$ pb.
\label{tab:cdftt} }       
\begin{tabular}{llll}
\hline\noalign{\smallskip}
analysis & SM non $\mathrm{t \bar{t}} $ &  $\mathrm{t \bar{t}} $ & data  \\
\noalign{\smallskip}\hline\noalign{\smallskip}
dilepton & $2.7 \pm 0.7$ & $10.9 \pm 1.4$ & 13 \\
lepton+jets(1 tag) & $21.8 \pm 3.0$ & $54.0 \pm 4.3$ & 49 \\
lepton+jets(2 tags) & $1.3 \pm 0.3$ & $10 \pm 1$ & 8 \\
lepton+tau & $1.3 \pm 0.2$ & $2.3 \pm 0.3$ & 2 \\
\noalign{\smallskip}\hline
\end{tabular}
\end{table}
For high tan~$\beta$, the charged Higgs decay mostly in $\bar \tau \nu_\tau$\cite{CDF:Hpp}.
The same procedure using the parameter \newline $\mathrm{Br(t \to H^+ b)}$ instead of tan~$\beta$
leads to an upper limit on $\mathrm{Br(t \to H^+ b)}$ as a function of the charged Higgs
mass shown on figure~\ref{fig:ttresult2} which is comparable to the one shown on 
figure~\ref{fig:cdfHtaunulimit}.

\section{Doubly charged Higgs}
\label{sec:HiggsPP}
The supersymmetric version of Left-Right symmetric models predicts the existence
of two light doubly charged Higgs\cite{Gunion:1989in}. One, $\mathrm{H^{++}_L}$,
 couples only to 
left-handed fermions and the other, $\mathrm{H^{++}_R}$, only to right-handed fermions.
At the TeVatron, the main production channel is the pair production of 
$\mathrm{H^{++}H^{--}}$ through a $\gamma/\mathrm{Z^0}$ s-channel exchange. The resulting
cross section depends essentially on the $\mathrm{H^{++}}$ mass and electroweak
quantum numbers\cite{Muhlleitner:2003me}. For all the results presented in this section, lower 
limits on the Higss mass are derived from upper limits on the 
$\mathrm{p \bar p \to H^{++}H^{--} + X}$ cross section using the cross section 
computed in~\cite{Muhlleitner:2003me} as a function of the Higgs mass.

\subsection{$\mathrm{H^{++}\to \mu^+ \mu^+}$}
\label{subsec:Hmumu}
\subsubsection{{D\O} analysis}
\label{subsec:HmumuD0}
\begin{figure}
\includegraphics[width=0.45\textwidth,height=0.35\textwidth,angle=0]{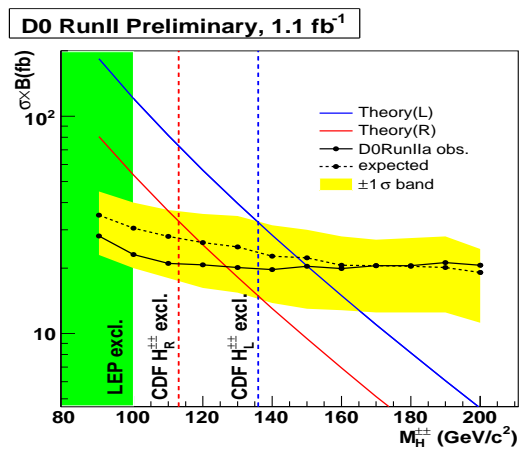}
\caption{Observed and expected 95 \% confidence level upper limit on the cross section 
of the process $\mathrm{p \bar p \to H^{++}H^{--} + X}$ as a function of the 
$\mathrm{H^{++}}$ mass assuming a 100\% decay rate for $\mathrm{H^{++}\to\mu^+\mu^+}$.
\label{fig:D0Hmumu}}       
\end{figure}

The {D\O} collaboration reports a search for the process \newline
$\mathrm{p \bar p \to H^{++}H^{--} + X \to \mu^+\mu^+\mu^-\mu^- + X}$ using a data 
sample of 1.1~fb$^{-1}$ \cite{D0:5458}. The analysis requests the 
events to be selected by a di-muon trigger. Each muon used in the event should
be isolated in the tracker and in the calorimeter and have a transverse momentum
of at least 15~GeV/c. Muon pairs compatible with being due to the passage of a cosmic muon 
are rejected. The events should contain at least one pair of like sign muons 
having a difference in azimuth less than 2.5 radians and an invariant mass greater than
30~GeV/c$^2$. The event should contain at least three identified muons. 
The analysis leads to 3 events observed for an 
expected background of $3.1 \pm 0.5$. Assuming a 100\% branching ratio of 
doubly charged Higgs into a muon pair, a 95\% confidence level upper limit
on the $\mathrm{p \bar p \to H^{++}H^{--} + X}$ cross section is derived as 
shown on figure~\ref{fig:D0Hmumu}. 
This translates into lower mass limits of 126.5~GeV/c$^2$ for the $\mathrm{H^{++}_R}$ and
150~GeV/c$^2$ for the $\mathrm{H^{++}_L}$.

\subsubsection{CDF analysis}
\label{subsec:HmumuCDF}
The CDF collaboration reports a similar analysis as the one described in 
section~\ref{subsec:HmumuD0} but for a data sample of 
242~pb$^{-1}$\cite{Acosta:2004uj}. In this analysis, events are selected if they 
fire a single muon trigger. The event should contain a pair of like sign
isolated muons with an invariant mass greater than 80~GeV/c$^2$. Pairs compatible with 
being due to a cosmic muon are rejected. In total, this analysis selects no event for
an expectation of $0.8 \pm 0.5$. Assuming  a 100\% decay rate for 
$\mathrm{H^{++}\to\mu^+\mu^+}$,  95\% confidence level lower limits on mass are derived. 
Those limits, shown on figure~\ref{fig:D0Hmumu}
are 113~GeV/c$^2$ for the $\mathrm{H^{++}_R}$ and 136~GeV/c$^2$ for the $\mathrm{H^{++}_L}$.

\subsection{other $\mathrm{H^{++}}$ decays}
\label{subsec:otherHpp}
Other decays looked for cover $\mathrm{H^{++}\to e^+ e^+}$, $\mathrm{H^{++}\to}$ 
$\mathrm{e^+ \mu^+}$,
$\mathrm{H^{++}\to e^+ \tau^+}$ and $\mathrm{H^{++}\to \mu^+ \tau^+}$. 

The like sign dielectron signal is searched for in 235~pb$^{-1}$ of CDF data where 
events are selected if they fire a di-electromagnetic trigger and if they contain 
a pair of like sign isolated electron with a tranverse momentum above 30~GeV/c. 
The invariant mass of the di-electron pair should be greater than 100~GeV/c$^2$. 
No event is found for an expectation of $1.1 \pm 0.4$ \cite{Acosta:2004uj}.
Corresponding lower mass limits for the doubly charged Higgs are displayed 
in table~\ref{tab:Hppsummary}.

The like sign electron-muon signal is looked for in 240 ~pb$^{-1}$ of CDF data
where events are selected if they fire a single electromagnetic trigger and 
contain a like sign pair made of an isolated electron and an isolated muon. The
invariant mass of the pair should be greater than 80~GeV/c$^2$. No event is found
for an expectation of $0.4 \pm 0.2$ \cite{Acosta:2004uj}. Corresponding lower mass limits 
for the doubly charged Higgs are displayed in table~\ref{tab:Hppsummary}.
 
Like sign electron+tau lepton and muon+tau lepton is searched for in the CDF data.
For the electron+tau lepton (resp. muon+tau lepton) analysis, it is required to have 
one isolated electron (resp. muon) with a transverse momentum above 20~GeV/c, a ha\-dro\-nic
decaying tau lepton or an electron with a transverse momentum above 15~GeV/c and 
a hadronic decaying tau lepton or an electron (resp. or an electron or a muon) 
with a transverse momentum above 10~GeV/c. The data samples are then divided 
according to the presence or absence of a fourth isolated high transverse momentum
lepton. Extra selections based on di-lepton masses, missing transverse energy, the 
scalar sum of missing transverse momentum and transverse momenta of all leptons are 
performed as described in \cite{CDF:etaumutau}. No event is observed in data for 
an expectation of $0.24 \pm 0.27$ (resp. $0.04 \pm 0.05$) for the three 
(resp. four) leptons eletron+tau lepton sample and of $0.27 \pm 0.125$ 
(resp. $0.14 \pm 0.05$) for the three (resp. four) leptons muon+tau lepton sample.
Data sample luminosity and observed lower mass limits are shown in 
table~\ref{tab:Hppsummary}.

Table~\ref{tab:Hppsummary} summarizes all the current TeVatron lower mass limits for the
doubly charged Higgs boson for various decays and also for a search for Higgs as 
heavy charged particle decaying after the tracker\cite{Acosta:2005np}.

%
\begin{table}
\caption{95\% confidence level lower mass limits on the doubly charged Higgs.
\label{tab:Hppsummary} }      
\begin{tabular}{lllll}
\hline\noalign{\smallskip}
hypothesis & exp & luminosity & \multicolumn{2}{l}{mass limit}  \\
 & &  & \multicolumn{2}{l}{(GeV/c$^2$)}  \\
 & &  &  $\mathrm{H^{++}_L}$ &  $\mathrm{H^{++}_R}$   \\
\noalign{\smallskip}\hline\noalign{\smallskip}
$\mathrm{Br(H^{++}\to e^+ e^+)}=1$ \cite{Acosta:2004uj}& CDF & 235~pb$^{-1}$ & 133 &\\
$\mathrm{Br(H^{++}\to \mu^+ \mu^+)}=1$ \cite{Acosta:2004uj}& CDF & 242~pb$^{-1}$ & 136 & 113\\
$\mathrm{Br(H^{++}\to \mu^+ \mu^+)}=1$ \cite{D0:5458}& {D\O} & 1.1~fb$^{-1}$ & 150 & 126.5\\
$\mathrm{Br(H^{++}\to e^+ \mu^+)}=1$ \cite{Acosta:2004uj}& CDF & 240~pb$^{-1}$ & 115 &\\
$\mathrm{Br(H^{++}\to e^+ \tau^+)}=1$ \cite{CDF:etaumutau}& CDF & 350~pb$^{-1}$ & 113.6 &\\
$\mathrm{Br(H^{++}\to \mu^+ \tau^+)}=1$ \cite{CDF:etaumutau}& CDF & 322~pb$^{-1}$ & 112.1 &\\
Long lived\cite{Acosta:2005np} & CDF & 292~pb$^{-1}$ & 133 & 109 \\
\noalign{\smallskip}\hline
\end{tabular}
\end{table}

\subsection*{Acknowledgments}
The CDF and {D\O} collaborations wish to thank the staffs at Fermilab and
at their collaborating institutions.

%
 \bibliographystyle{unsrt}
 \bibliography{geraldgrenier.bib}

\begin{thebibliography}{10}

\bibitem{Nilles:1983ge}
Hans~Peter Nilles.
\newblock Supersymmetry, supergravity and particle physics.
\newblock {\em Phys. Rept.}, 110:1, 1984.

\bibitem{Gunion:1989in}
J.~F. Gunion, J.~Grifols, A.~Mendez, B.~Kayser, and Fredrick~I. Olness.
\newblock Higgs bosons in left-right symmetric models.
\newblock {\em Phys. Rev.}, D40:1546, 1989.

\bibitem{Carena:2002es}
Marcela~S. Carena and Howard~E. Haber.
\newblock Higgs boson theory and phenomenology. ((v)).
\newblock {\em Prog. Part. Nucl. Phys.}, 50:63--152, 2003.

\bibitem{CDF:8353}
The~CdF Collaboration.
\newblock Search for anomalous tau production in b-tagged top quark events.
\newblock CdF preliminary results, CDF/ANA/EXOTIC/PUBLIC/8353, June 2006.

\bibitem{CDF:Hpp}
The~CdF Collaboration.
\newblock A search of charged higgs in the decay products of pair-produced top
  quarks.
\newblock CdF preliminary results,
  http://www-cdf.fnal.gov/physics/new/top/2005/ljets/
  charged\_higgs/higgs/V2/HiggsAnalysis\_publicV2.html, 2005.

\bibitem{Acosta:2004uw}
D.~Acosta et~al.
\newblock Measurement of the $t\bar{t}$ production cross section in $p\bar{p}$
  collisions at $\sqrt{s} = 1.96$ tev using dilepton events.
\newblock {\em Phys. Rev. Lett.}, 93:142001, 2004.

\bibitem{Acosta:2004hw}
D.~Acosta et~al.
\newblock Measurement of the $t\bar{t}$ production cross section in $p\bar{p}$
  collisions at $\sqrt{s} = 1.96$ tev using lepton + jets events with secondary
  vertex $b-$tagging.
\newblock {\em Phys. Rev.}, D71:052003, 2005.

\bibitem{Abulencia:2005et}
A.~Abulencia et~al.
\newblock A search for $t \to \tau \nu q$ in $t\bar{t}$ production.
\newblock {\em Phys. Lett.}, B639:172, 2006.

\bibitem{Muhlleitner:2003me}
Margarete Muhlleitner and Michael Spira.
\newblock A note on doubly-charged higgs pair production at hadron colliders.
\newblock {\em Phys. Rev.}, D68:117701, 2003.

\bibitem{D0:5458}
The~{D\O} Collaboration.
\newblock Search for pair production of doubly-charged higgs bosons in the
  h$^{++}$h$^{--}\to 4\mu$ final state at {D\O}.
\newblock {D\O} preliminary results, conference note, {D\O} Note 5458-CONF,
  August 2007.

\bibitem{Acosta:2004uj}
D.~Acosta et~al.
\newblock Search for doubly-charged higgs bosons decaying to dileptons in
  $p\bar{p}$ collisions at $\sqrt{s} = 1.96$ tev.
\newblock {\em Phys. Rev. Lett.}, 93:221802, 2004.

\bibitem{CDF:etaumutau}
The~CdF Collaboration.
\newblock Search for doubly charged higgs in lepton flavor violating decays
  involving taus.
\newblock CdF preliminary results,
  http://www-cdf.fnal.gov/physics/exotic/r2a/20060406.HPlusPlus/, April 2006.

\bibitem{Acosta:2005np}
D.~Acosta et~al.
\newblock Search for long-lived doubly-charged higgs bosons in $p\bar{p}$
  collisions at $\sqrt{s} = 1.96$ tev.
\newblock {\em Phys. Rev. Lett.}, 95:071801, 2005.

\end{thebibliography}
%
%
%

\end{document}